\documentclass[aps,prc,reprint,groupedaddress]{revtex4-1}
\usepackage{graphicx}
\def\vc#1{\mbox{\boldmath $#1$}}
\def\Be{{^{8}{\rm Be}}}
\def\C{{^{12}{\rm C}}}
\def\Ox{{^{16}{\rm O}}}

\begin{document}


\title{Monopole excitation of the Hoyle state and linear-chain state in $\C$}


\author{Y.~\textsc{Funaki}}
\affiliation{School of Physics and Nuclear Energy Engineering and IRCNPC, Beihang University, Beijing 100191, China
}


\date{\today}

\begin{abstract}
\begin{description}
\item[Background] Two new $J^\pi=0^+$ states are recently observed around a few MeV above the Hoyle state (the second $0^+$ state in $\C$). The characters of them are only poorly discussed in theory and are still mysterious.
\item[Purpose] I give for the first time a comprehensive understanding of the structures of the $0^+$ states by analyzing their wave functions, and discuss relationship with the Hoyle state, similarities, and differences between the states.
\item[Method] I extend a microscopic $\alpha$-cluster model called Tohsaki-Horiuchi-Schuck-R\"opke (THSR) wave function so as to incorporate $2\alpha+\alpha$ asymmetric configuration explicitly. The so-called $r^2$-constraint method to effectively eliminate spurious continuum components is also used.
\item[Results] The $0_3^+$ state is shown to have a very large squared overlap with a single configuration of the extended THSR wave function in an orthogonal space to the Hoyle state as well as to the ground state. The $0_4^+$ state has a maximal squared overlap with a single extended THSR wave function with an extremely prolately-deformed shape.
\item[Conclusions] The $0_3^+$ state appears as a family of the Hoyle state to have a higher nodal structure in the internal motions of the $3\alpha$ clusters, due to the orthogonalization to the Hoyle state. The $0_4^+$ state dominantly has a linear-chain structure, where the $3\alpha$ clusters move freely in a non-localized way, like a one-dimensional gas of the $3\alpha$ clusters.
\end{description}

\end{abstract}


\maketitle

\section{Introduction}

Nuclear cluster structure in the Hoyle state, the second $J^\pi=0^+$ state at $7.65$ MeV in $\C$, has been discussed for a long time by many authors~\cite{Ho74,Ka78,Ue77,Fu80,De87,Pi97,Su02,Ya04,Ya05,Ku05,Fu06a,Ar06,Ch07,En07,Oh13,Fu13,Fr14}. Cluster model approaches play an important role in understanding the structure and clarified that it has well developed $3\alpha$ cluster structure with more dilute density than that of saturation in the ground state, where $\alpha$ clusters weakly interact with each other in a relative $S$-wave~\cite{Ho74,Ka78,Ue77,Fu80,De87}. In the recent fifteen years, however, the understanding of the Hoyle state structure has been much deepened, by an advent of a new type of microscopic cluster model wave function, which is referred to as the Tohsaki-Horiuchi-Schuck-R\"opke (THSR) wave function~\cite{To01,Fu02,Fu15a}. This wave function retains a structure that constituent $\alpha$ clusters are loosely bound like a gas and occupy an identical orbit, and this structured phenomenon is now called the $\alpha$ condensation. One of the most important properties for the THSR wave function is to give a single and optimal configuration that is equivalent to a solution of full microscopic three-body problem~\cite{Fu03,Fu05,Fu09}. Since it is well known that the solutions of the full microscopic three-body problem via $3\alpha$ Resonating Group Method (RGM)~\cite{Ka78} and via Generator Coordinate Method (GCM)~\cite{Ue77} nicely reproduce many experimental data for the Hoyle state, like the energy, width, electro-magnetic properties, etc, the equivalence leads to the conclusion that the Hoyle state exists as the $\alpha$ condensate composed of a weakly interacting and gaslike $3\alpha$ clusters. 

Not only the Hoyle state but also some other excited states in $\C$ trigger a special interest in recent years. For example, the second $J^\pi=2^+$ state was theoretically predicted by the use of the cluster models almost 40 years ago~\cite{Ka78,Ue77,De87}, but it was very recently when the existence was confirmed in several experiments~\cite{Fr09,It11,Fy11,Zi11,Zi13}, with a pioneering work by Itoh {\it et al.} as a beginning~\cite{It04}. Besides the second $2^+$ state, a new $4^+$ state is also observed at $13.3$ MeV recently~\cite{Fr11}. The new $2^+$ and $4^+$ states are now considered to form a rotational family with the Hoyle state, though the detailed rotational structure is under question~\cite{Ma14,Fu15}. While in Ref.~\cite{Ma14} a simple rotational structure based on a triangular shape of the $3\alpha$ clusters is assumed, it is pointed out in Ref.~\cite{Fu15} that this is not simply considered to be an ordinary rotational band that lies on $J(J+1)$ line, due to the $\alpha$-condensate nature of the Hoyle state, where the third $0^+$ state $(0_3^+)$ above the Hoyle state also plays an important role.

A new experimental information is also given for the famous broad $0^+$ state observed at $10.3$ MeV with a width of $2.7$ MeV~\cite{Jo03,Fy03,It04,Di05}. Itoh {\it et al.} pointed out that the broad $0^+$ peak is decomposed into two peaks, giving the $0_3^+$ and $0_4^+$ states at $1.77$ MeV and $3.29$ MeV above the $3\alpha$ threshold, with the widths of $1.45$ MeV and $1.42$ MeV, respectively~\cite{It13}. They also found a decay property that the $0_4^+$ state dominantly decays into $\Be(2^+)+\alpha(D)$ while the $0_3^+$ state decays into $\Be(0^+)+\alpha(S)$. Some theoretical studies consistently reproduce the resonance parameters, where semi-microscopic~\cite{Ku05,Oh13} or non-microscopic~\cite{Ri11,Ish14} $3\alpha$ model is adopted. In particular, in Ref.~\cite{Ku05}, the authors applied the Complex Scaling Method (CSM) and Analytic Continuation of Coupling Constant (ACCC) method to the $3\alpha$ Orthogonality Condition Model (OCM). They suggested by extrapolation that the $0_3^+$ state has an $S$-wave dominant structure with more dilute density than that of the Hoyle state. The observed decay property and resonance parameters of the $0_3^+$ and $0_4^+$ states are also reproduced by the recent calculation by the present author using an extended version of the THSR wave function, where $\Be+\alpha$ correlation can be taken into account~\cite{Fu15}. He further showed more directly by using the THSR wave function that the $0_3^+$ state is given rise to as a result of the monopole excitation from the Hoyle state to have dominantly a higher nodal structure, where the $\alpha$ cluster orbits around the $\Be(0^+)$ in an $S$-wave with four nodes. He also showed that the $0_4^+$ state has the largest $S^2$-factor in the channel of the $\alpha$ cluster coupling with $\Be(2^+)$ in a $D$-wave. This is consistent with the previous result of the $3\alpha$ GCM calculation~\cite{Ue77}, where a large Reduced Width Amplitude (RWA) from the $\Be(2^+)+\alpha(D)$ channel is obtained, although their calculation fails in reproducing the observed $0_3^+$ state. I should also mention that the Antisymmetrized Molecular Dynamics (AMD) and Fermionic Molecular Dynamics (FMD) calculations reproduce the $0_4^+$ state and predict that the state dominantly has an intrinsic configuration of bent-armed shape of the $3\alpha$ clusters, like a linear-chain structure that was originally proposed by Morinaga~\cite{Mo56}, although the observed $0_3^+$ state is also missing in their calculations~\cite{Ne04,Ch07,En07}. In Ref.~\cite{Ka86}, the linear-chain component of the $0_4^+$ state obtained by the $3\alpha$ OCM is identified to be $(\lambda,0)$ configuration in the Elliott $SU(3)$ model, which is calculated to be about $56$ \%. 

In this paper, I investigate the excited $J^\pi=0^+$ states obtained in the previous work~\cite{Fu15} by using the extended THSR wave function and the so-called $r^2$-constraint method~\cite{Fu05,Fu06b,Fu10}. I focus on how much components concentrate on a single configuration of the extended THSR wave function with deformation parameters, and discuss physical natures of the states. In Sec.~\ref{sec:2}, the original THSR wave function is explained, and then as its natural extension the extended THSR wave function is introduced. In Sec.~\ref{sec:3}, the structures of the $0_2^+$, $0_3^+$, and $0_4^+$ states are discussed. Squared-overlap surfaces between the states and single configurations of the THSR wave function in their deformation parameter space are calculated. Sec.~\ref{sec:4} is devoted to conclusion.

\section{THSR wave function} \label{sec:2}

The original THSR wave function with deformation~\cite{Fu02} is described below,
\begin{eqnarray}
&&\Phi^{\rm THSR}(\vc{\beta}) \nonumber \\
&&= {\cal A}\Big[\prod_{i=1}^3 \exp \Big\{ - 2\hspace{-0.2cm} \sum_{k=x,y,z}\hspace{-0.2cm} \frac{(R_{ik}-X_{k})^2}{b^2+2\beta_{k}^2}\Big\} \phi(\alpha_i) \Big], \nonumber \\ 
&& = {\cal A}\Big[ \exp \Big\{ - \sum_{i=1}^2 \mu_i \hspace{-0.2cm} \sum_{k=x,y,z}\hspace{-0.1cm} \frac{\xi_{ik}^2}{b^2+2\beta_{k}^2}\Big\} \phi(\alpha_1)\phi(\alpha_2)\phi(\alpha_3) \Big],\nonumber \\
\label{eq:thsr} 
\end{eqnarray}
with ${\cal A}$ being the antisymmetrization operator acting on the 12 nucleons, $\phi(\alpha_i)$ the internal wave function of the $i$-th $\alpha$ particle assuming a $(0s)^4$ configuration, like,
\begin{equation}
\phi(\alpha_i)\propto \exp\big[-\sum_{1\leq j<k \leq 4}(\vc{r}_{4(i-1)+j}-\vc{r}_{4(i-1)+k})^2/(8b^2)\big],
\end{equation}
$\vc{R}_i=\sum_{j=1}^{4}\vc{r}_{4(i-1)+j}/4$, $\vc{X}=\sum_{j=1}^{12}\vc{r}_j/12$, the position vectors of the $i$-th $\alpha$ particle and of total center-of-mass, respectively, $\vc{\xi}_1=\vc{R}_2-\vc{R}_1$ and $\vc{\xi}_2=\vc{R}_3-(\vc{R}_1+\vc{R}_2)/2$ the Jacobi coordinates between the $\alpha$ particles, and $\mu_i=i/(i+1)$. The parameters $b$ and $\vc{\beta}$ characterize the size of the constituent $\alpha$ particle, and the size and shape of the total nucleus, respectively, though the axial symmetry $\beta_x=\beta_y$ is assumed throughout this study.

The extended version of the THSR wave function which I utilize in this work is a natural extension of the original form, Eq.~(\ref{eq:thsr}), as follows:
\begin{eqnarray}
&&\Phi^{\rm THSR}(\vc{\beta}_1,\vc{\beta}_2) \nonumber \\
&&\hspace{-0.25cm} ={\cal N} {\cal A}\Big[ \exp \Big\{\hspace{-0.1cm} - \hspace{-0.1cm} \sum_{i=1}^2 \mu_i \hspace{-0.2cm} \sum_{k=x,y,z}\hspace{-0.2cm} \frac{\xi_{ik}^2}{b^2+2\beta_{ik}^2}\Big\} \phi(\alpha_1)\phi(\alpha_2)\phi(\alpha_3) \Big], \label{eq:2} \nonumber \\
\end{eqnarray}
where the single parameter $\vc{\beta}$ is decomposed into $\vc{\beta}_1$ and $\vc{\beta}_2$ corresponding to the two Jacobi coordinates, $\vc{\xi}_1$ and $\vc{\xi}_2$, and ${\cal N}$ is a normalization constant. This allows us to include $(\Be + \alpha)$-type configuration beyond the original THSR wave function, where all $\alpha$ clusters are restricted to move in an identical orbit. 

This model wave function provides a picture that constituent clusters of a nucleus are trapped into a potential without any geometrical rigid configuration of the clusters under the constraint of antisymmetrization. The center-of-mass wave functions of the constituent clusters are assumed to have deformable Gaussian shapes with widths that are variational parameters and characterize the spatial size and shape of the nucleus. This is mentioned as a ``container'' picture or non-localized concept of cluster structures in some recent publications (see Ref.~\cite{Fu15a} and references therein). Not only the gaslike cluster states like the Hoyle state $(3\alpha)$ and $\Be$ $(2\alpha)$ but also ordinary cluster states, which had been believed to have non-gaslike localized cluster structures, such as the inversion doublet band with $\Ox+\alpha$ structure~\cite{Zh12}, $3\alpha$ and $4\alpha$ linear-chain structure~\cite{Su14}, $2\alpha+\Lambda$ structure in ${^9_\Lambda {\rm Be}}$~\cite{Fu14}, $2\alpha+n$ and $2\alpha+2n$ structures in ${^9{\rm Be}}$~\cite{Ly15} and ${^{10}{\rm Be}}$~\cite{Ly16}, respectively, etc, can all be described by this THSR ansatz with almost $100$ \% accuracy.

Since the excited states above the Hoyle state were observed as resonances with non-negligible widths, it is more likely that the bound state approximation does not work well for those states. I therefore use a technique to effectively eliminate continuum components that get mixed with the resonances, so-called $r^2$-constraint method, which is also used in Refs.~\cite{Fu05,Fu06b,Fu10,Fu15} and the effectiveness is already guaranteed. In this technique, by considering the fact that in calculations of bound states, pseudo continuum states are shown to have large root mean square (r.m.s.) radii, compared to those of resonances and bound states, one can remove effectively the spurious continuum components in the following way: First I solve the following equation,
\begin{eqnarray}
&&\hspace{-0.4cm} \sum_{\vc{\beta}_1^\prime,\vc{\beta}_2^\prime} \langle \Phi^{\rm THSR}_{J=0}(\vc{\beta}_1,\vc{\beta}_2) |{\widehat {\cal O}}_{\rm rms}- \{R^{(\gamma)}\}^2 |\Phi^{\rm THSR}_{J=0}(\vc{\beta}_1^\prime,\vc{\beta}_2^\prime) \rangle \nonumber \\ 
&& \times g^{(\gamma)}(\vc{\beta}_1^\prime,\vc{\beta}_2^\prime)=0, \label{eq:cutoff1}
\end{eqnarray}
with ${\widehat {\cal O}}_{\rm rms} = \sum_{i=1}^{12}(\vc{r}_i-\vc{X})^2/12$, and $\Phi^{\rm THSR}_{J=0}(\vc{\beta}_1,\vc{\beta}_2)={\widehat P}_{J=0} \Phi^{\rm THSR}(\vc{\beta}_1,\vc{\beta}_2)$, where ${\widehat P}_{J=0}$ is the angular-momentum projection operator onto $J=0$. The eigenstates in Eq.~(\ref{eq:cutoff1}) can be written as,
\begin{equation}
\Phi^{(\gamma)}_{J=0} = \sum_{\vc{\beta}_1,\vc{\beta}_2}g^{(\gamma)}(\vc{\beta}_1,\vc{\beta}_2) \Phi^{\rm THSR}_{J=0}(\vc{\beta}_1,\vc{\beta}_2). \label{eq:cutoff2}
\end{equation}
Next I adopt, as bases to diagonalize Hamiltonian, the eigenstates belonging to eigenvalues satisfying $R^{(\gamma)} \le R_{\rm cut}$ in Eq.~(\ref{eq:cutoff2}), as follows: 
\begin{equation}
\sum_{\gamma^\prime} \langle \Phi^{(\gamma)}_{J=0} |H| \Phi^{(\gamma^\prime)}_{J=0} \rangle f_{\lambda}^{(\gamma^\prime)} =E_\lambda f_{\lambda}^{(\gamma)}, \label{eq:eigenwf}
\end{equation}
and obtain the eigenfunction,
\begin{equation}
\Psi^{(\lambda)}_{J=0} = \sum_{\gamma} f_{\lambda}^{(\gamma)} \Phi^{(\gamma)}_{J=0}. \label{eq:wf}
\end{equation}
For Hamiltonian, the same effective nucleon-nucleon interaction as used in Ref.~\cite{Fu15}, Volkov No.~2 force~\cite{Vol65}, is adopted, where the strength parameters are slightly modified~\cite{Fu80}. The cutoff radius is now taken to be $R_{\rm cut}=6.0$ fm as in Ref.~\cite{Fu15}.

\section{Results and Discussions}\label{sec:3}

\begin{table}[htbp]
\begin{center}
\caption{Energies, widths, r.m.s. radii, and monopole transition strengths. Units of energies and widths are MeV, of radii ${\rm fm}$, $M(E0)$ ${\rm fm}^2$. Experimental data (Exp.), THSR, OCM1, OCM2, FMC, and AMD are taken from Refs.~\cite{It11}, ~\cite{Fu15}, ~\cite{Ku05}, ~\cite{Oh13}, ~\cite{Ch07}, ~\cite{En07}, respectively.}
\begin{tabular}{clcccccc}
\hline\hline
 &  & Exp. & THSR & OCM1 & OCM2 & FMD & AMD \\
\hline
\multicolumn{2}{c}{$E(0_2^+)-E_{3\alpha}$} & \multicolumn{1}{c}{$0.38$} & $0.23$ & $0.76$ & $0.75$ & $0.4$ & $\sim 3.5$ \\
\multicolumn{2}{c}{$E(0_3^+)-E_{3\alpha}$} & \multicolumn{1}{c}{$1.77(9)$} & $2.6$ & $1.66$ & $0.79$ &  &  \\
\multicolumn{2}{c}{$E(0_4^+)-E_{3\alpha}$} & \multicolumn{1}{c}{$3.29(6)$} & $3.9$ & $4.58$ & $4.59$ & $2.85$ & $\sim 6.5$ \\
\multicolumn{2}{r}{$\Gamma(0_2^+)$ $(\times 10^{-6})$} & \multicolumn{1}{c}{$8.5$} & $7.6$ & $240$ & $880$ &  & $40$ \\
\multicolumn{2}{c}{$\Gamma(0_3^+)$} & \multicolumn{1}{c}{$1.45(18)$} & $1.1$ & $1.48$ & $1.68$ &  &  \\
\multicolumn{2}{c}{$\Gamma(0_4^+)$} & \multicolumn{1}{c}{$1.42(8)$} & $0.58$ & $1.1$ & $1.0$ &  & $0.4$ \\
\multicolumn{2}{c}{$R_{\rm rms}(0_2^+)$} &  & $3.7$ & $4.23$ &  & $3.38$ & $3.27$ \\
\multicolumn{2}{c}{$R_{\rm rms}(0_3^+)$} &  & $4.7$ &  &  &  &  \\
\multicolumn{2}{c}{$R_{\rm rms}(0_4^+)$} &  & $4.2$ & $3.49$ &  & $4.62$ & $3.98$ \\
\multicolumn{2}{c}{$M(E0, 0_2^+ \rightarrow 0_1^+)$} & \multicolumn{1}{c}{$5.4(2)$} & $6.3$ &  &  & $6.53$ & $6.7$ \\
\multicolumn{2}{c}{$M(E0, 0_3^+ \rightarrow 0_2^+)$} &  & $\sim 35$ &  &  &  &  \\
\multicolumn{2}{c}{$M(E0, 0_4^+ \rightarrow 0_2^+)$} &  & $\sim 1.0$ &  &  &  & $2.0$ \\
\hline\hline
\end{tabular}\label{tab:1}
\end{center}
\end{table}

In TABLE~\ref{tab:1}, the calculated observables, energy, $\alpha$-decay width, monopole transition strengths are compared with the experimental data and those of the other calculations. The present calculation consistently reproduces the corresponding experimental data. The r.m.s. radius of the $0_3^+$ state is the largest of the $0^+$ states and the much stronger monopole transition strength between the $0_2^+$ and $0_3^+$ states, than that between the $0_2^+$ and $0_1^+$ state (see Ref.~\cite{Ya08}), can be seen. These indicate, as discussed in Ref.~\cite{Fu15}, that the $0_3^+$ state is excited by monopole transition from the Hoyle state to have a higher nodal structure between $\Be$ and $\alpha$ cluster. 

As I mentioned in Sec.~\ref{sec:2}, I adopted Volkov No. 2 force as an effective nucleon-nucleon
interaction in this work. I also checked if the results depend on the choice of the force. For example, when
I adopt Volkov No. 1 force, the energies of the $0_2^+$, $0_3^+$, and $0_4^+$ states become slightly higher 
by about $1$ MeV, i.e. $E-E_{3\alpha}= 1.2$ MeV, $4.2$ MeV, and $5.0$ MeV, respectively.
However, I confirmed that qualitative features of these states that will be discussed 
in this section, such as the behaviours of squared overlaps, do not change at all. 

I also mention that only the THSR ansatz lists rather complete physical quantities, including the r.m.s. radii and monopole transition strengths, since the wave functions of both $0_3^+$ and $0_4^+$ states are definitely obtained. As mentioned in Introduction, the AMD and FMD calculations cannot reproduce the $0_3^+$ state, and in OCM$+$CSM$+$ACCC calculation in Ref.~\cite{Ku05}, denoted as OCM1 in this table, the wave function of the $0_3^+$ state cannot be obtained.

\begin{figure}[htbp]
\begin{center}
\includegraphics[scale=0.9]{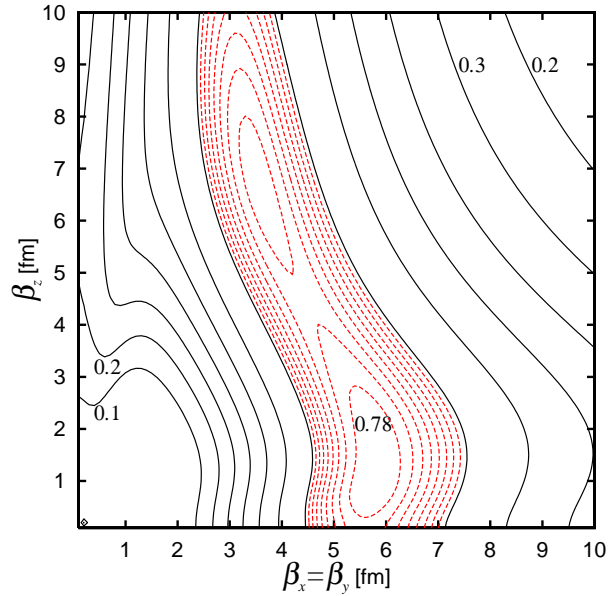}
\caption{(color online) Contour map of the squared overlap ${\cal O}_{\lambda}(\vc{\beta}_1=\vc{\beta}_2)$ in Eq.~(\ref{eq:ovlp1}) with $\lambda=2$, i.e. for the $0_2^+$ state, in two parameter space, $\vc{\beta}_1=\vc{\beta}_2=(\beta_x=\beta_y,\ \beta_z)$. Black solid curves are drawn in a step of $0.1$ and red dotted curves, which cover the region of ${\cal O}_{\lambda}(\vc{\beta}_1=\vc{\beta}_2)\ge 0.81$, are in a step of $0.01$.}
\label{fig:1}
\end{center}
\end{figure}

\begin{figure}[htbp]
\begin{center}
\includegraphics[scale=0.9]{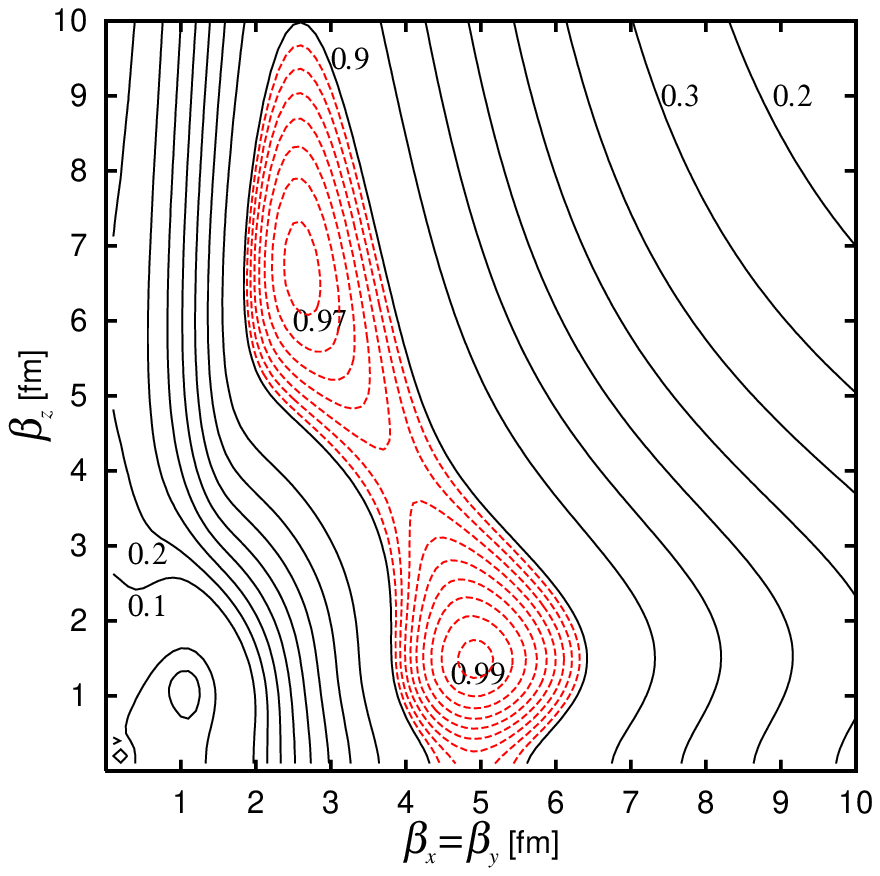}
\caption{(color online) Contour map of the squared overlap ${\widetilde {\cal O}}_{\lambda}(\vc{\beta}_1=\vc{\beta}_2)$ in Eq.~(\ref{eq:ovlp2}) with $\lambda=2$, i.e. for the $0_2^+$ state, in two parameter space, $\vc{\beta}_1=\vc{\beta}_2=(\beta_x=\beta_y,\ \beta_z)$. Black solid curves are drawn in a step of $0.1$ and red dotted curves, which cover the region of ${\widetilde {\cal O}}_{\lambda}(\vc{\beta}_1=\vc{\beta}_2)\ge 0.91$, are in a step of $0.01$.}
\label{fig:2}
\end{center}
\end{figure}

In Ref.~\cite{Fu15}, the author investigated the Hoyle band $(0_2^+,2_2^+,4_2^+)$ and the $0_3^+$ and $0_4^+$ states by using the extended THSR wave function. All the states obtained in his calculations are derived as solutions of the Hill-Wheeler equation Eq.~(\ref{eq:wf}), where many bases of THSR-type configurations are superposed. This will, however, make it unclear how the THSR picture mentioned above is realized for those states, since the superposition of the THSR configurations might break its original picture. Thus, in order to investigate how the THSR picture holds for the $0^+$ states, I calculate the following two quantities: 
\begin{equation}
{\cal O}_\lambda(\vc{\beta}_1,\vc{\beta}_2) = |\langle \Phi_{J=0}^{\rm THSR}(\vc{\beta}_1,\vc{\beta}_2) | \Psi^{(\lambda)}_{J=0} \rangle |^2 \label{eq:ovlp1}
\end{equation}
and 
\begin{equation}
{\widetilde {\cal O}}_\lambda(\vc{\beta}_1,\vc{\beta}_2) = |\langle {\widetilde N}_\lambda {\widehat P}_{\lambda} \Phi_{J=0}^{\rm THSR}(\vc{\beta}_1,\vc{\beta}_2) | \Psi^{(\lambda)}_{J=0} \rangle |^2. \label{eq:ovlp2}
\end{equation}
The former is the squared overlap of the $0_2^+$ $(\lambda=2)$, $0_3^+$ $(\lambda=3)$ and $0_4^+$ $(\lambda=4)$ states in Eq.~(\ref{eq:wf}) with the single configuration of the angular-momentum-projected THSR wave function, $\Phi_{J=0}^{\rm THSR}(\vc{\beta}_1,\vc{\beta}_2)$. The latter is also the squared overlap of the $0^+$ states in Eq.~(\ref{eq:wf}) with the single configuration of the angular-momentum-projected THSR wave function but the lower $0^+$ states components are subtracted by the projection operator defined as, ${\widehat P}_\lambda \equiv 1-\sum_{i=1}^{\lambda-1}|\Psi^{(\lambda)}_{J=0} \rangle \langle \Psi^{(\lambda)}_{J=0} |$, from the single THSR wave function. ${\widetilde N}_\lambda$ is a normalization constant of the wave function ${\widehat P}_\lambda \Phi_{J=0}^{\rm THSR}(\vc{\beta}_1,\vc{\beta}_2)$.

In FIG.~\ref{fig:1}, I show the contour map of the squared overlap ${\cal O}_{\lambda=2}(\beta_x=\beta_y,\ \beta_z)$ in two parameter space $\beta_x=\beta_y$ and $\beta_z$ with $\vc{\beta}=\vc{\beta}_1=\vc{\beta}_2$. One can see that maximum value of the squared overlap amounts to $0.79$ at $\beta_x=\beta_y= 5.8$ fm and $\beta_z= 1.1$ fm. While this large value indicates that the Hoyle state is expressed by a single configuration of the THSR wave function with $\vc{\beta}_1=\vc{\beta}_2$, the Hoyle state is orthogonal to the ground state, and therefore the orthogonality condition should be imposed on the THSR wave function to describe the Hoyle state. I then calculate the second quantity of the squared overlap discussed above, ${\widetilde {\cal O}}_{\lambda=2}(\beta_x=\beta_y,\ \beta_z)$ with $\vc{\beta}=\vc{\beta}_1=\vc{\beta}_2$. Figure~\ref{fig:2} shows the contour map in two parameter space $\vc{\beta}_1=\vc{\beta}_2=(\beta_x=\beta_y,\ \beta_z)$. The maximal value is found to be $0.992$ at $(\beta_x=\beta_y,\ \beta_z)=(5.0\ {\rm fm},\ 1.5\ {\rm fm})$, which is surprisingly large value, almost $100$ \%. This large value was already found before in Refs.~\cite{Fu03,Fu09} with the THSR ansatz of $\vc{\beta}_1=\vc{\beta}_2$, but I should note that in the present case this large value is also obtained in the calculation with larger model space $\vc{\beta}_1 \ne \vc{\beta}_2$, where $2\alpha$ and $\alpha$ asymmetric configuration is allowed for. The fact that nevertheless this large value is again obtained strongly suggests that in the Hoyle state the $3\alpha$ particles, democratically, without a strong $\Be + \alpha$ correlation, take an identical motion, so that the $3\alpha$ condensate state is realized. One can also find that the region denoted by dotted curve in red, where the squared overlap is more than $0.91$, is widely ranged from prolately $\beta_x=\beta_y < \beta_z$ to oblately $\beta_x=\beta_y > \beta_z$ deformed shapes, passing through the spherical one $\beta_x=\beta_y = \beta_z$, with large $\beta_x=\beta_y,\ \beta_z$ values. This supports the idea that the Hoyle state does not have any definite intrinsic shape but have a gaslike configuration of the $3\alpha$ particles. 

I should also note that the projection operator ${\widehat P}_{\lambda=2}$, which removes the ground state component with compact structure, plays a role as a repulsive force to prevent the $3\alpha$ clusters from being resolved and to form well developed $3\alpha$ cluster structure for the Hoyle state, due to its orthogonality condition. This is essentially the same as the situation of $\Be$, where the antisymmetrization operator ${\cal A}$ removes the Pauli forbidden states, to construct a structural repulsive core between the two $\alpha$ clusters.

\begin{figure}[htbp]
\begin{center}
\includegraphics[scale=0.9]{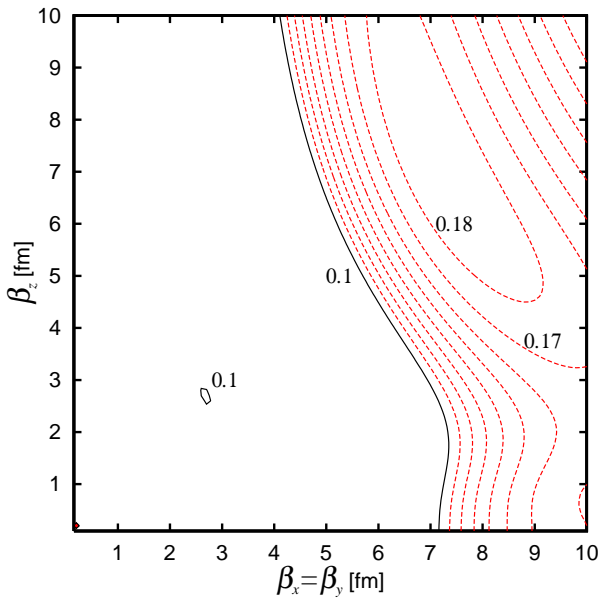}
\caption{(color online) Contour map of the squared overlap ${\cal O}_{\lambda}(\vc{\beta}_1=\vc{\beta}_2)$ in Eq.~(\ref{eq:ovlp1}) with $\lambda=3$, i.e. for the $0_3^+$ state, in two parameter space, $\vc{\beta}_1=\vc{\beta}_2=(\beta_x=\beta_y,\ \beta_z)$. Black solid curves are drawn in a step of $0.1$ and red dotted curves, which cover the region of ${\cal O}_{\lambda}(\vc{\beta}_1=\vc{\beta}_2)\ge 0.11$, are in a step of $0.01$.}
\label{fig:3}
\end{center}
\end{figure}

\begin{figure}[htbp]
\begin{center}
\includegraphics[scale=0.9]{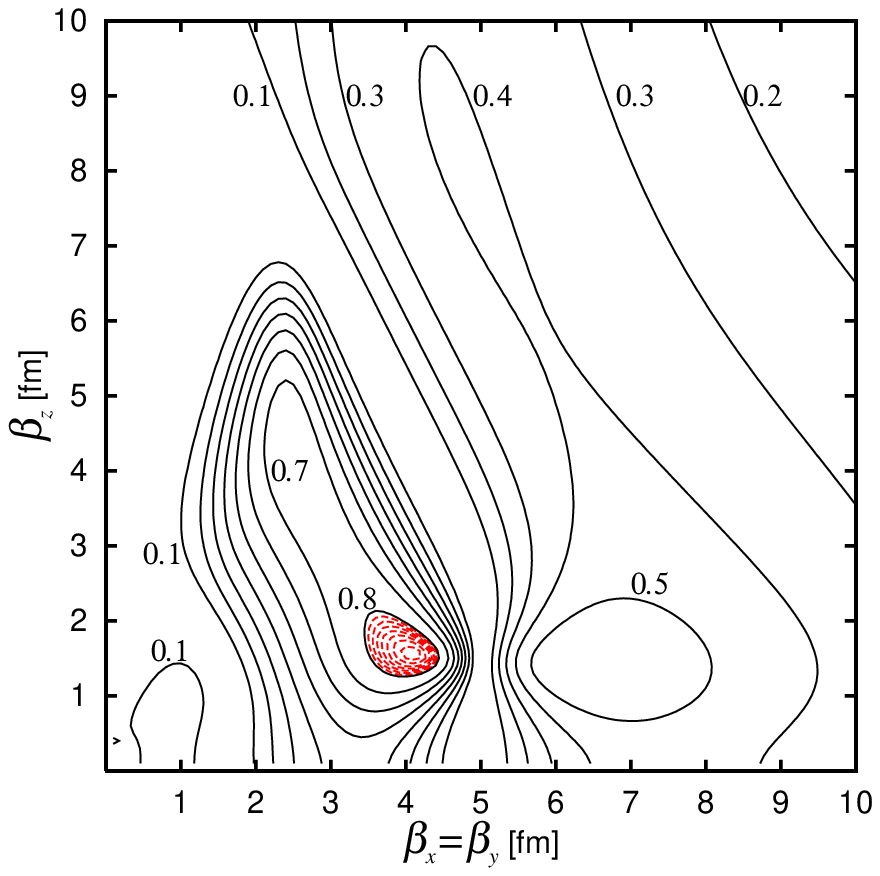}
\caption{(color online) Contour map of the squared overlap ${\widetilde {\cal O}}_{\lambda}(\vc{\beta}_1=\vc{\beta}_2)$ in Eq.~(\ref{eq:ovlp2}) with $\lambda=3$, i.e. for the $0_3^+$ state, in two parameter space, $\vc{\beta}_1=\vc{\beta}_2=(\beta_x=\beta_y,\ \beta_z)$. Black solid curves are drawn in a step of $0.1$ and red dotted curves, which cover the region of ${\widetilde {\cal O}}_{\lambda}(\vc{\beta}_1=\vc{\beta}_2)\ge 0.81$, are in a step of $0.01$.}
\label{fig:4}
\end{center}
\end{figure}

\begin{figure}[htbp]
\begin{center}
\includegraphics[scale=0.9]{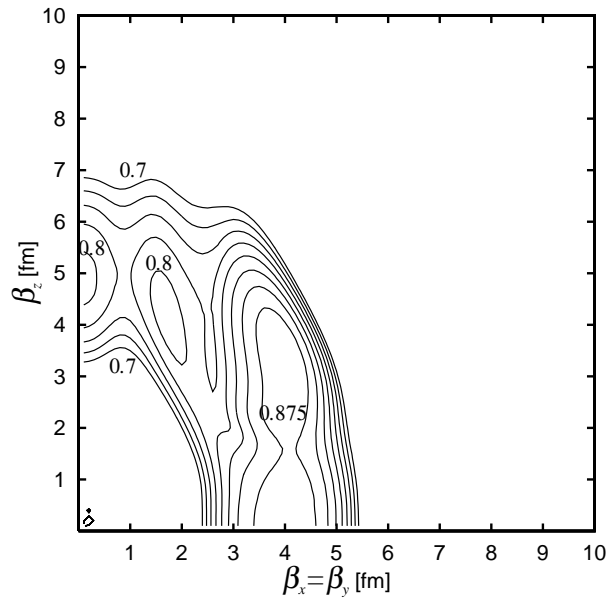}
\caption{Contour map of the squared overlap ${\widetilde {\cal O}}_{\lambda}(\vc{\beta}_1,\vc{\beta}_2)$ in Eq.~(\ref{eq:ovlp2}) with $\lambda=3$, i.e. for the $0_3^+$ state, in two parameter space, $\vc{\beta}_1=(\beta_x=\beta_y,\ \beta_z)$, where the maximal values of ${\widetilde {\cal O}}_{\lambda}(\vc{\beta}_1,\vc{\beta}_2)$ obtained by varying $\vc{\beta}_2$ are shown. The region of ${\widetilde {\cal O}}_{\lambda}(\vc{\beta}_1,\vc{\beta}_2)\ge 0.7$ is only shown in a step of $0.025$ of the contour lines.}
\label{fig:5}
\end{center}
\end{figure}

Next I show the contour map of the squared overlap for the $0_3^+$ state, ${\cal O}_{\lambda=3}(\beta_x=\beta_y,\ \beta_z)$ in FIG.~\ref{fig:3} and ${\widetilde {\cal O}}_{\lambda=3}(\beta_x=\beta_y,\ \beta_z)$ in FIG.~\ref{fig:4} in two parameter space $\beta_x=\beta_y$ and $\beta_z$. In FIG.~\ref{fig:3}, the contour lines more than $0.1$ are denoted by dotted ones in red, in a step of $0.01$. One can see that the $0_3^+$ state does not have any squared overlap that is more than 0.1 in internal region of $\beta_x=\beta_y$ and $\beta_z$. It only has at most $0.19$ in an outside region around $\beta_x=\beta_y=7$ fm, $\beta_z=7$ fm. I also check the squared overlap in the model space of $\vc{\beta}_1 \neq \vc{\beta}_2$, i.e. ${\cal O}_{\lambda=3}(\vc{\beta}_1,\vc{\beta}_2)$ in four parameter space, but no large squared overlap more than 0.19 can be obtained.

However, this situation changes drastically when I consider the THSR model space of its single configuration that is orthogonal to the Hoyle state as well as to the ground state, i.e. ${\widehat P}_{\lambda=3}\Phi_{J=0}^{\rm THSR}$, where the projection operator is defined as ${\widehat P}_{\lambda=3}=1-|\Psi^{(\lambda=1)}_{J=0} \rangle \langle \Psi^{(\lambda=1)}_{J=0} |-|\Psi^{(\lambda=2)}_{J=0} \rangle \langle \Psi^{(\lambda=2)}_{J=0} |$. Then in FIG.~\ref{fig:4}, I show the squared overlap ${\widetilde {\cal O}}_{\lambda=3}(\beta_x=\beta_y,\ \beta_z)$ in two parameter space, $\vc{\beta}_1=\vc{\beta}_2=(\beta_x=\beta_y,\ \beta_z)$. The contour lines more than $0.8$ are denoted by dotted ones in red, in a step of $0.01$. One can see that contrary to FIG.~\ref{fig:3}, very large squared overlap appears in the internal region. The largest one amounts to $0.89$ at $(\beta_x=\beta_y,\ \beta_z)=(4.1\ {\rm fm},\ 1.6\ {\rm fm})$. One can also see that this large value quickly decreases toward around $(\beta_x=\beta_y,\ \beta_z)=(5.0\ {\rm fm},\ 1.5\ {\rm fm})$, which corresponds to the point giving the maximum squared overlap for the Hoyle state in FIG.~\ref{fig:2}. This is of course due to the effect of the orthogonalization operator ${\widehat P}_{\lambda=3}$. These results clearly indicate that in order to describe the $0_3^+$ state, the orthogonalization to the Hoyle state as well as to the ground state is the most essential. This implies that the $0_3^+$ state is intimately related to the Hoyle state, like its family. Due to this orthogonalization, the $0_3^+$ state is considered to exist as an excitation mode from the Hoyle state with respect to the internal motions of the $3\alpha$ clusters. 

In the previous paper Ref.~\cite{Fu15}, for the $0_3^+$ state, the large RWA of $\Be(0^+) + \alpha(S)$ component with a higher node than in the Hoyle state is calculated. Also as shown in TABLE~\ref{tab:1}, the much larger monopole strength from the Hoyle state than a single particle strength, $\sim 35\ {\rm fm}^2$, is obtained. From these results, it was concluded that the $0_3^+$ state is given rise to as a result of the strong monopole transition from the Hoyle state, to have a higher nodal structure. These results are consistent with the present results, giving the interpretation that the monopole transition or vibration mode is brought about by the orthogonalization to the Hoyle state, to necessarily provide the higher nodes in the internal motions of the $3\alpha$ clusters.

In order to further investigate how $0_3^+$ state contains the $\Be+\alpha$ correlation, I then calculate the squared overlap in four parameter space, ${\widetilde {\cal O}}_{\lambda=3}(\vc{\beta}_1,\ \vc{\beta}_2)$. For a given $\vc{\beta}_1$ value, the maximal squared overlap obtained by varying $\vc{\beta}_2$ value is shown in the contour map of FIG.~\ref{fig:5}, as a function of two parameters $\vc{\beta}_1=((\vc{\beta}_1)_x=(\vc{\beta}_1)_y,\ (\vc{\beta}_1)_z)$. The region giving more than the squared overlap of $0.7$ is only shown in FIG.~\ref{fig:5} with contour lines in a step of $0.025$. While the largest value appears at oblately deformed region, like FIG.~\ref{fig:4}, it is also found that there appear two maxima in the prolately deformed region, which are about $0.81$, at $\vc{\beta}_1=((\vc{\beta}_1)_x=(\vc{\beta}_1)_y,\ (\vc{\beta}_1)_z)=(0.1\ {\rm fm},\ 4.8\ {\rm fm})$ and $\vc{\beta}_1=((\vc{\beta}_1)_x=(\vc{\beta}_1)_y,\ (\vc{\beta}_1)_z)=(1.7\ {\rm fm},\ 4.3\ {\rm fm})$. The corresponding $\vc{\beta}_2$ values are $\vc{\beta}_2=((\vc{\beta}_2)_x=(\vc{\beta}_2)_y,\ (\vc{\beta}_2)_z)=(3.4\ {\rm fm},\ 4.3\ {\rm fm})$ and $\vc{\beta}_2=((\vc{\beta}_2)_x=(\vc{\beta_2})_y,\ (\vc{\beta}_2)_z)=(3.5\ {\rm fm},\ 4.0\ {\rm fm})$, respectively. Since $\vc{\beta}_1$ and $\vc{\beta}_2$ correspond to the deformation parameters of $\Be$ and the remaining $\alpha$-cluster motion, respectively, and $\Be$ is expressed to have prolately deformed value for $\vc{\beta}_1$, this large squared overlap values indicate that the $0_3^+$ state has sizable $\Be+\alpha$ correlation, which is also consistent with the previous calculation of the $\Be+\alpha$ RWA.

\begin{figure}[htbp]
\begin{center}
\includegraphics[scale=0.9]{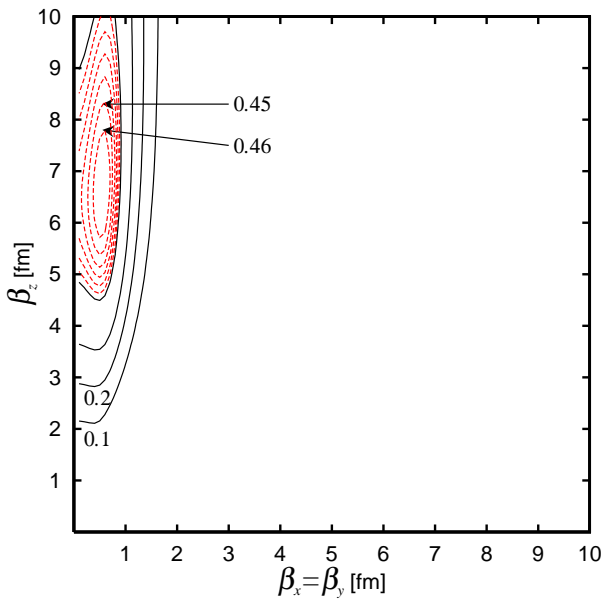}
\caption{(color online) Contour map of the squared overlap ${\cal O}_{\lambda}(\vc{\beta}_1=\vc{\beta}_2)$ in Eq.~(\ref{eq:ovlp1}) with $\lambda=4$, i.e. for the $0_4^+$ state, in two parameter space, $\vc{\beta}_1=\vc{\beta}_2=(\beta_x=\beta_y,\ \beta_z)$. Black solid curves are drawn in a step of $0.1$ and red dotted curves, which cover the region of ${\cal O}_{\lambda}(\vc{\beta}_1=\vc{\beta}_2)\ge 0.41$, are in a step of $0.01$.}
\label{fig:6}
\end{center}
\end{figure}

\begin{figure}[htbp]
\begin{center}
\includegraphics[scale=0.7, angle=270]{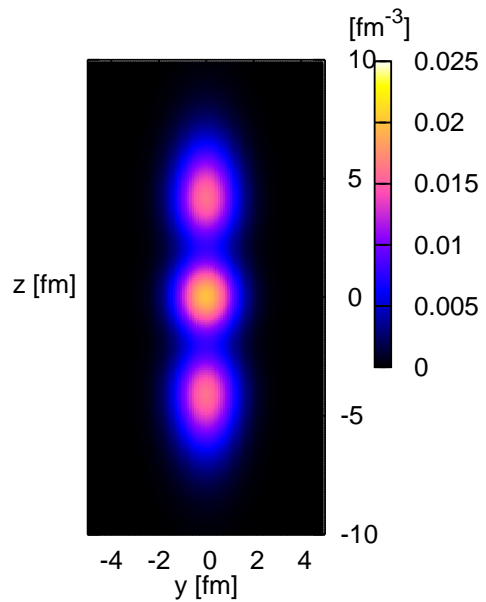}
\caption{(color online) Intrinsic density profile generated from the THSR wave function before angular-momentum projection, with $\vc{\beta}_1=\vc{\beta}_2=(\beta_x=\beta_y,\ \beta_z)=(0.6\ {\rm fm},\ 6.7\ {\rm fm})$, which gives the maximal squared overlap, $0.47$, in FIG.~\ref{fig:6}.}
\label{fig:7}
\end{center}
\end{figure}

Finally I discuss the structure of the $0_4^+$ state. As I mentioned in Introduction, in the AMD, FMD, and $3\alpha$ GCM calculations, the $0_3^+$ state may be missing and only the $0_4^+$ state was obtained. The dominant intrinsic configuration of the $0_4^+$ state in the AMD and FMD calculations shows a bent-armed structure of the $3\alpha$ clusters, resembling the linear-chain structure. On the other hand, in the OCM$+$CSM$+$ACCC and $3\alpha$ GCM calculations, the $0_4^+$ state has a structure that the $\alpha$ cluster predominantly couples in a $D$-wave with the $\Be(2^+)$ core. The same result is also given by the RWA analysis in the present ansatz in Ref.~\cite{Fu15}. The problem is whether the interpretation of the linear-chain-like structure for the $0_4^+$ state is reasonable or not, since the AMD and FMD calculations cannot reproduce the $0_3^+$ and $0_4^+$ state simultaneously. Actually it was recently reported by Suhara {\it et al.}, that the orthogonality condition to the lower states plays an important role for the survival of the linear-chain structure state in $\C$~\cite{Su15}.

In FIG.~\ref{fig:6}, I show the contour map of the squared overlap for the $0_4^+$ state, ${\cal O}_{\lambda=4}(\vc\beta_x=\beta_y,\ \beta_z)$, in two parameter space, $\vc{\beta}_1=\vc{\beta}_2=(\beta_x=\beta_y,\ \beta_z)$. The contour lines giving more than $0.4$ are denoted by dot in red, in a step of $0.01$. This contour map has a characteristic feature, where the strongly prolate deformation is only allowed to have a non-negligible squared overlap amplitude. Except for this prolately deformed region, the squared overlap is less than $0.1$. The largest value is $0.47$, which is not so much large but clearly indicates the $3\alpha$ linear-chain structure. I can consider this situation as follows: It is shown that the extremely prolately-deformed shaped THSR wave function has very small overlap with the other shaped THSR wave function (see FIG.~2 in Ref.~\cite{Fu05}). Since the configuration space other than the extremely prolately-deformed region is already used by the Hoyle state and the $0_3^+$ state (see FIG.~\ref{fig:1} and FIG.~\ref{fig:3}), as well as by the ground state for the more compact region, the $0_4^+$ state has no choice but using the remaining configuration, to result in having the extremely prolately-deformed shape, i.e. linear-chain structure. I also mention that this feature of the $0_4^+$ state is quite different from the behaviour of the Hoyle state, in which the $\vc{\beta}$ parameter space giving the large squared overlap is widely spanned as shown in FIG.~\ref{fig:1}. While the feature for the Hoyle state implies that any definite intrinsic wave function for this state is difficult to be uniquely determined, the feature of the $0_4^+$ state allows for the definite intrinsic shape.

The parameter values giving the maximal squared overlap is calculated to be $\vc{\beta}_1=\vc{\beta}_2=(0.6\ {\rm fm},\ 6.7\ {\rm fm})$, which is close to $\vc{\beta}_1=\vc{\beta}_2=(0.1\ {\rm fm},\ 5.1\ {\rm fm})$ that was obtained in a rather ideal one-dimension situation in Ref.~\cite{Su14}. In Ref.~\cite{Su14}, it is discussed that largely prolately-deformed THSR wave function shows one-dimensional $\alpha$ condensate of $3\alpha$ clusters, which is fairly different from the ordinary picture of the linear-chain state with rigid-body $3\alpha$-cluster configuration arranged in a line in a spatially localized way. Thus I can say that the present $0_4^+$ state has the one-dimensional $\alpha$ condensate structure by around $50$ \%, where the $3\alpha$ clusters are loosely trapped into a prolately deformed potential like a one-dimensional gas. 

\begin{figure}[htbp]
\begin{center}
\includegraphics[scale=0.9]{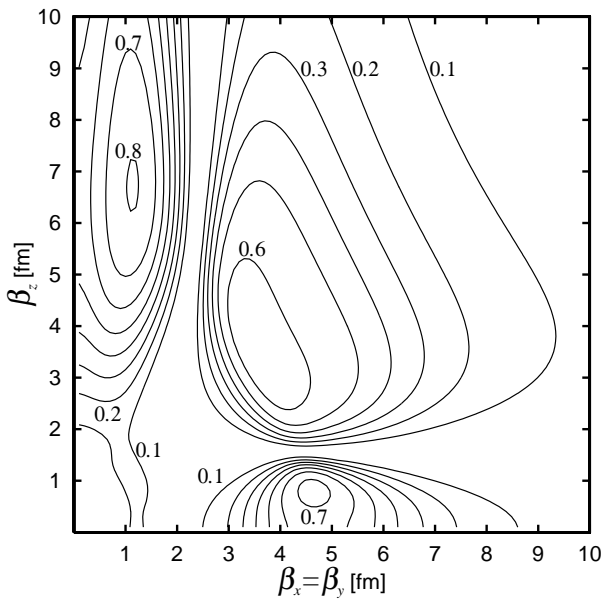}
\caption{Contour map of the squared overlap ${\widetilde {\cal O}}_{\lambda}(\vc{\beta}_1=\vc{\beta}_2)$ in Eq.~(\ref{eq:ovlp2}) with $\lambda=4$, i.e. for the $0_4^+$ state, in two parameter space, $\vc{\beta}_1=\vc{\beta}_2=(\beta_x=\beta_y,\ \beta_z)$, in a step of $0.1$.}
\label{fig:8}
\end{center}
\end{figure}

Figure.~\ref{fig:7} shows the intrinsic density profile generated from the single THSR wave function, before angular-momentum projection, with $\vc{\beta}_1=\vc{\beta}_2=(0.6\ {\rm fm},\ 6.7\ {\rm fm})$, which gives the maximal squared overlap in FIG.~\ref{fig:6}. This density distribution shows a clear linear-chain structure of the $3\alpha$ clusters, i.e. localized $\alpha$ clusters, with an extended long tail along the $z$-direction. As is discussed in Ref.~\cite{Su14}, this comes from the inter-$\alpha$ Pauli repulsion, as a kinematical effect, which makes this object look like keeping localized clustering. However, dynamics prefers a one-dimensional gas, according to the potential picture and the character of the THSR wave function mentioned above. This is particularly expressed as the long tail elongated along the $z$-direction in this figure.

Finally I show in FIG.~\ref{fig:8} the contour map in the orthogonal space, ${\widetilde {\cal O}}_{\lambda=4}(\beta_x=\beta_y,\ \beta_z)$ as usual, in two parameter space $\vc{\beta}_1=\vc{\beta}_2=(\beta_x=\beta_y,\ \beta_z)$. The largest squared overlap value increases up to $0.81$ and the $\vc{\beta}_1=\vc{\beta}_2$ value to give the maximum slightly moves toward spherical region, which is $(1.1\ {\rm fm},\ 6.6\ {\rm fm})$, but still shows very strongly prolately-deformed shape. On the one hand, this large value again supports that the $0_4^+$ state has a linear-chain shape. On the other hand, the second and third maxima also appear at $(4.6\ {\rm fm},\ 0.8\ {\rm fm})$ and $(3.5\ {\rm fm},\ 3.6\ {\rm fm})$, where the maximal values are $0.71$ and $0.68$, respectively. The former and the latter correspond to the oblately deformed shape and the spherical shape, respectively. This may suggest that due to the orthogonalization operator ${\widehat P}_{\lambda=4}$, some other correlations to unstabilize the linear-chain structure, like a bending mode, which allows for a spherical shape at a certain probability, take part in, as is discussed in Ref.~\cite{Su15}. 

\section{Conclusion}\label{sec:4}

In conclusion, I investigated the excited $J^\pi=0^+$ states in $\C$ by using the extended THSR wave function with $r^2$-constraint method. In particular, I focused on the $0_3^+$ and $0_4^+$ states, which were recently found in experiment. The physical properties of the states, relationship with the Hoyle state, and similarities and differences between them were discussed by calculating the squared overlap with the single configurations of the extended THSR wave function. The $0_3^+$ state was found to appear as a result of the orthogonalization to the Hoyle state as well as to the ground state, so that the strong monopole transition or vibrational transition is induced. The state is considered to be a family of the Hoyle state with a higher nodal structure in internal motions of the $3\alpha$ clusters. The $0_4^+$ state was shown to have a linear-chain structure as a dominant configuration, where the $3\alpha$ clusters move rather freely in a much elongated way along the $z$-axis, i.e. like a one-dimensional gas, though the density distribution shows a localized $3\alpha$ linear-chain structure, due to the inter-$\alpha$ Pauli repulsion. Besides the linear-chain configuration, some other correlations like a bending mode also seem to be mixed.

\section*{Acknowledgements}
The author wishes to thank B. Zhou, H. Horiuchi, T. Suhara, A. Tohsaki, G. R\"opke, P. Schuck, T. Yamada for many helpful discussions. This work is financially supported by HPCI project and JSPS KAKENHI Grant Number 25400288. The support of Beihang University under the ``Zhuoyue 100 Talents'' program is gratefully acknowledged.

\end{document}